\documentclass[12pt]{spieman}

\usepackage[final]{graphicx}
\usepackage{times,bbm,amsmath,amssymb}
\usepackage{epsfig,color}
\usepackage{xcolor}
\usepackage{hyperref}
\hypersetup{
    colorlinks = true
}
\usepackage{cleveref}

\usepackage[caption = false]{subfig}

\usepackage[greek,english]{babel}
\usepackage{thumbpdf,enumerate}
\usepackage{booktabs}
\usepackage{sidecap}
\usepackage[scaled=.8]{couriers}    
\usepackage{pstricks}
\usepackage{multirow}
\usepackage{placeins}
\usepackage{relsize}  
\usepackage{pst-grad,bm}
\usepackage{epigraph}
\usepackage{textcomp}
\usepackage{gensymb}
\usepackage{longtable}
\usepackage{booktabs}
\usepackage{gensymb}

\usepackage{soul}
\usepackage{ulem} 
\usepackage{acronym}

\usepackage{physics}

\usepackage{tikz}

\newcommand{\add}[1]{\textcolor{blue}{#1}}

\usepackage{bbold}
\usepackage{microtype}
\usepackage{listings}

\title{Characterization of multi-mode linear optical networks}

\author[a]{Francesco Hoch}
\author[a]{Taira Giordani}
\author[a]{Nicol\`o Spagnolo}
\author[b,c]{Andrea Crespi}
\author[c]{Roberto Osellame}
\author[a,*]{Fabio Sciarrino}

\affil[a]{Dipartimento di Fisica, Sapienza Universit\`{a} di Roma, Piazzale Aldo Moro 5, I-00185 Roma, Italy}
\affil[b]{Dipartimento di Fisica, Politecnico di Milano, Piazza Leonardo da Vinci, 32, I-20133 Milano, Italy}
\affil[c]{Istituto di Fotonica e Nanotecnologie, Consiglio Nazionale delle Ricerche (IFN-CNR), Piazza Leonardo da Vinci, 32, I-20133 Milano, Italy}

\newcommand{\comment}[1]{}

\begin{document}
\maketitle

\begin{abstract}
Multi-mode optical interferometers represent the most viable platforms for the successful implementation of several quantum information schemes that take advantage of optical processing. Examples range from quantum communication, sensing and computation, including optical neural networks, optical reservoir computing or simulation of complex physical systems. The realization of such routines requires high levels of control and tunability of the parameters that define the operations carried out by the device. This requirement becomes particularly crucial in light of recent technological improvements in integrated photonic technologies, which enable the implementation of progressively larger circuits embedding a considerable amount of tunable parameters. In this work, we formulate efficient procedures for the characterization of optical circuits in the presence of imperfections that typically occur in physical experiments, such as unbalanced losses and phase instabilities in the input and output collection stages. The algorithm aims at reconstructing the transfer matrix that represents the optical interferometer without making any strong assumptions about its internal structure and encoding. We show the viability of this approach in an experimentally relevant scenario, defined by a tunable integrated photonic circuit, and we demonstrate the effectiveness and robustness of our method. Our findings can find application in a wide range of optical setups, based both on bulk and integrated configurations.
\end{abstract}

\keywords{integrated photonics, device characterization, quantum networks}

{\noindent \footnotesize\textbf{*}Fabio Sciarrino,  \linkable{fabio.sciarrino@uniroma1.it} }

\section*{Introduction}

Linear optical networks are fundamental elements in several protocols for computation, communication and sensing. Recently, many schemes for computation have found their natural implementations through optical processing, such as neuromorphic and reservoir computing \cite{Vandoorne2014,Sande_2019_book, Gigan_reservoir, Nakajima2021}, optical neural networks \cite{Shen2017,Wang2022} and optical simulation \cite{Pietrangeli_2019, Okawachi2020, Leonetti_2021}. Large-scale photonic platforms are one of the most promising candidates to implement quantum information and quantum computation protocols \cite{Flamini_2018, Wang2020}.
Indeed, they have been extensively employed for quantum walks routines \cite{Gr_fe_2016, Cardano_2016, Esposito2022}, quantum machine learning algorithms \cite{Wang2017,Saggio2021,Spagnolo2022}, and, recently, for experiments that aim at demonstrating quantum advantage with photons \cite{Zhong_GBS_supremacy,Zhong_phase_2021, Madsen2022}. All these protocols in classical and quantum optics require complex interferometric structures composed of numerous optical components. Integrated photonics is one of the best candidates to realize such optical protocols in compact devices, offering in addition the capability to reconfigure the circuit operation \cite{Wang2020,Pelucchi2022}. The latest examples of multi-mode optical networks \cite{quix_20modes, Hoch2021} have shown a significant increase in the network complexity as well as in the number of control parameters.

The mathematical tool to model any linear optical processing in such experiments is the unitary matrix $U$ that describes the relation between input/output complex amplitudes of the electromagnetic field. Several relevant aspects connected to optical network programming have been identified. They range from engineering the optical setup for implementation of a given $U$, to the identification of universal architectures that can perform any transformation \cite{Reck_1994, Clements:16}.
Here we focus on the characterization of linear optical circuits, which requires a systematic methodology to verify the operation of the optical circuit, or more generally, to reconstruct the unitary matrix implemented by the network. This task turns out to be a non-trivial one in certain scenarios. For example, interferometers based on a bulk optic implementation can be typically decomposed in elementary units that can be addressed and characterized separately. On the contrary, this procedure is typically not viable for integrated photonic circuits, and the network needs to be analysed as a single element.

\begin{figure*}[t]
    \centering
    \includegraphics[width=0.99\textwidth]{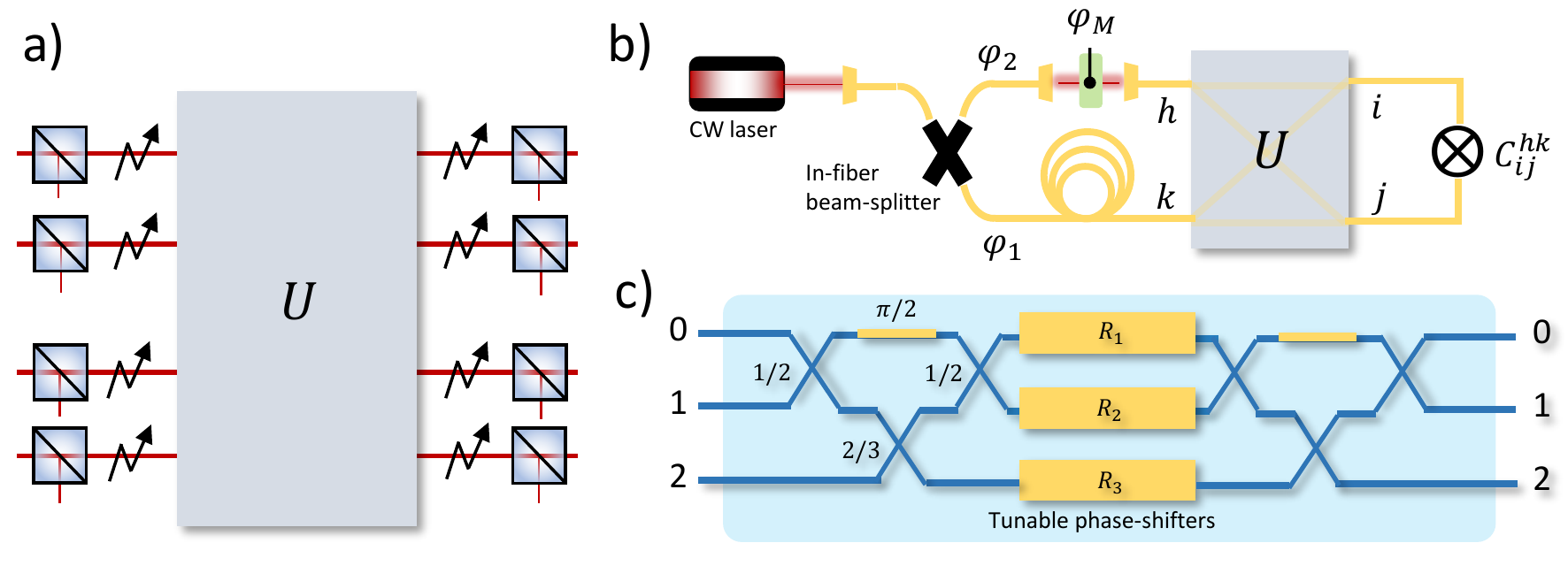}
    \caption{\textbf{Reconstruction of multi-mode optical circuits}. a)The model of a multi-mode interferometer considered in this work. It is composed by the \add{ideal} optical circuit described by the unitary transformation $U$ plus layers of mode-dependent losses at the input and at the output (represented by beam splitters in the figure), and phase instabilities (represented by sparks). Output losses take into account also possible differences in the detection efficiencies among the modes. b) Scheme for the measurement of second-order cross-correlations $C^{hk}_{ij}$ with coherent light emitted by a continuous wave laser (CW). The latter is coupled in single-mode fiber and split into two beams by an in-fiber beam splitter. The two beams enter the interferometer in modes $h,k$. The phase modulation $\varphi_M$ performed by a liquid crystal compensates the fiber phase fluctuations $\varphi=\varphi_1 -\varphi_2$ to satisfy the conditions in Eq. \eqref{eq:conditions}. c) The 3-mode integrated chip employed to test the reconstruction algorithm. It is composed of a sequence of two tritter structures. Each tritter comprises three beam-splitters, whose reflectivity is reported in the figure, and a phase-shift equal to $\pi/2$. Between the two tritters there are three heaters $\{R_1,\, R_2,\, R_3\}$ that dynamically control the optical phases between the two structures via the thermo-optic effect.
    }
    \label{fig:chip}
\end{figure*}

Several techniques have been developed to achieve full characterization of integrated photonic devices by exploiting different degrees of knowledge of the network's internal structure and various measurement approaches with classical or quantum light. 
These algorithms can be divided into two main groups. The first one is called \emph{black box approach} and exploits only the information provided by the measurements, without assumptions on the internal structure of the optical network. The second one is the \emph{white box approach}, which exploits knowledge of the optical network structure together with the information obtained from a set of appropriately chosen measurements.
White box approaches usually make use of an optimization algorithm to estimate the parameters of the given architecture \cite{Tillmann_2016, Spagnolo2017}.
It is clear that this family of reconstruction algorithms strongly depends on the knowledge of the relation between the optical component parameters and the matrix elements, and thus requires accurate modelling of the system including noise processes.
On the contrary, black box approaches aim at characterizing a multi-mode linear optical interferometer without using any information about its structure. They are very useful when the structure is inaccessible, untrusted or too complex to be modelled \cite{Hoch2021}. A black-box method for linear optical circuits was proposed in Ref. \cite{laing2012superstable} and subsequently refined in Ref. \cite{Dhand_2016}. The authors presented an analytical algorithm to reconstruct the elements of the unitary matrix $U$ from quantum light measurement, via single- and two-photon experiments. The algorithm is advantageous since it permits to retrive the matrix $U$ even in the presence of losses and optical phase instabilities due to, for example, fiber connections in the input and output stages. However, such a method requires quantum light input states, is a slow procedure and the accuracy is limited by noisy experimental data. In fact, the result of this reconstruction method is typically employed as a starting point for further numerical optimization to improve the stability of the solution. 
An alternative method exploits only classical field intensities measurements \cite{Rahimi-Keshari:13}. The moduli of the matrix elements are measured from the field intensities distribution in the output modes, while, the complex phases are estimated by a measurement of the interference fringes between two coherent beams. 
The two procedures for the moduli and the phases estimations are independent and they can be mapped directly onto the unitary matrix without the need to apply any further optimization algorithms \cite{Rahimi-Keshari:13, Tillmann_2016}.
With this approach, a crucial requirement for a correct phase measurement is to perform the phase scan in times much shorter than the typical timescale of phase fluctuations of the system. In addition, it cannot be used in the presence of mode-dependent losses in the collection stages. Other black-box algorithms based on coherent light measurements always require high phase stability during the scan in the input and output sections \cite{Popoff2010_TM, Popoff_2011, suess2020rapid}, thus making them not viable for optical setups with in-fiber connections, which are nevertheless typical for integrated photonic devices. The last class we mention are machine-learning algorithms \cite{Cimini2021, Suprano_Zia_2021} that need large sets of data to learn the correct transformation. Very recently, Kuzmin et al. \cite{Kuzmin:21} simulated the application of a supervised-learning strategy for the calibration of a reconfigurable interferometer and experienced an unfavourable scaling of the training set size with the number of modes in the black box scenario.

In this work, we propose a new black box approach that aims at solving some open issues mentioned in the past algorithms. The goal is to provide a methodology to identify separately mode-dependent losses and matrix elements of $U$ via coherent light measurements. 
In particular, we first present two algorithms to estimate the amount of loss unbalancement in the injection and collection stages of a linear optical interferometer and consequently characterize the moduli of the elements of the unitary matrix. Then, we move to the problem of measuring the phases of the unitary matrix elements. Previous algorithms \cite{laing2012superstable, Dhand_2016} exploited second-order quantum optical correlations such as the Hong-Ou-Mandel effect \cite{Hong87}, or the first-order classical correlations in Mach–Zehnder like interferometric structures \cite{Rahimi-Keshari:13}. 
Since there are mathematical analogies between classical and quantum second-order correlations \cite{PhysRevA.81.023834}, we propose to replace such quantities with the second-order correlations of classical light in a Hanbury-Brown\&Twiss like experiment \cite{HANBURYBROWN1956}. This approach combines some advantages of both the previous methods, i.e. the simplicity in the use of classical light\cite{Rahimi-Keshari:13} and the independence from losses and optical phase instabilities, due to the fibers, 
characteristic of the methods that employ two-photon pairs as input states \cite{laing2012superstable, Dhand_2016}.
Moreover, we show that the proposed classical second-order measurements depend only on the matrix phases and not on the moduli as for the quantum correlations. This allows for completely independent estimations of the phases, input/output losses and the moduli of the unitary matrix. From an experimental point of view, this is an important improvement that reduces the propagation of the error along the characterization process.
Thus, this method can be applied in a general scenario and can be effective for different experimental platforms, ranging from bulk to integrated and in-fiber optical setups.

\section*{Overview on black-box linear optical circuits reconstruction}
Any ideal linear optical interferometer can be represented by a unitary matrix $U$, acting linearly on the annihilation (creation) operators of the electromagnetic field in the input modes $a_j$ and transforming to the annihilation (creation) operators in the output modes $b_i$
\begin{equation}
    b_{i} = \sum_j U_{ij} a_j
    \label{eq:unitary}
\end{equation}
where $U_{ij}$ are the elements of the unitary matrix. 
The same relation holds for classical states of light, by replacing the operators $a_i$ and $b_i$ with the complex field $E_i$ in Eq. \eqref{eq:unitary}. In other words, the elements of the unitary matrix $U_{ij}$ completely characterize the field amplitudes propagation through a multi-mode optical network. In general, the set of optical modes $i$ may represent any degree of freedom of light, such as polarization, path, time of arrival, frequency, angular and transverse momentum.

Optical losses deviate the interferometer operation from unitarity.
To take into account the losses we consider biased insertion losses at the input and at collection stages, and balanced internal losses which are known to commute with the unitary matrix \cite{Oszmaniec_2018}. This assumption is not the most general since it considers the optical transformation $U$ without any internal unbalances among the modes. Currently, most of the photonic circuits are designed in such a way that losses are practically unbiased along the evolution and can be factorized to one of the ends of the interferometer \cite{Clements:16}. Then, our model of losses can find applications in many scenarios. We model losses as in Fig.~\ref{fig:chip}a with a set of beam splitters placed on the input and output modes. We further consider the presence of unknown (and possibly unstable) phase terms on these modes. The device is thus described by a matrix $T = \mathcal{D}_1 U \mathcal{D}_2$, where $\mathcal{D}_1$ and $\mathcal{D}_2$ are diagonal matrices that account for such losses and additional phases.
The matrix elements of the unitary part are expressed as $U_{ij}=\tau_{ij} \text{e}^{\text{i} \phi_{ij}}$, where $\tau_{ij}$ and $\phi_{ij}$ are the matrix moduli and phases respectively. In what follows, we briefly analyse the two most general algorithms in the literature to reconstruct the matrix $U$. The first algorithm exploits quantum light \cite{laing2012superstable} and the second one is based on coherent light measurements \cite{Rahimi-Keshari:13}.

If one employs measurements with Fock states at the input of the interferometer and photon-number-resolving detection at the output, the results will be insensitive to the (unstable) phase terms at the input and at the output. This means that the matrix $U$ and all the matrices $U'$ in the form $U' = F_1 U F_2$, where $F_1$ and $F_2$ are a unitary diagonal matrix, are equivalent. Another invariance property of these measurements carried out with Fock states is that the outcomes do not change with respect to the conjugate operation $U' = U^*$. Given these equivalence relations, it is not necessary to reconstruct the actual unitary matrix implemented by the interferometer but only a representative element of its class of equivalence. This is commonly defined by choosing a matrix with real-valued elements in the first row and first column ($\phi_{0i} = 0$ and $\phi_{i0} = 0$) together with the condition that the element $U_{11}$ has positive phase ($\phi_{11} > 0$).
In Ref. \cite{laing2012superstable}, the authors presented an algorithm to reconstruct the value of moduli and phases of the representative unitary matrix elements via single-photon intensity and two-photon measurements, the latter via the visibility $\mathcal{V}$ of Hong-Ou-Mandel (HOM) interference \cite{Hong87}.
Labelling the input modes of the two photons as $h,k$, and the output ones as $i,j$, the visibility is defined as 
\begin{equation}
    \mathcal{V}^{hk}_{ij} = 1 - \frac{(P^{hk}_{ij})^{\text{I}}}{(P^{hk}_{ij})^{\text{D}}} = \frac{-2 \tau_{jk} \tau_{ih} \tau_{ik} \tau_{jh}}
    {\tau_{jk}^2 \tau_{ih}^2+ \tau_{ik}^2 \tau_{jh}^2 } \cos\left( \phi_{jk} + \phi_{ih}- \phi_{ik} -\phi_{jh}\right) 
\label{eq:vis}
\end{equation}
where $(P^{hk}_{ij})^{\text{I}}$ is the probability to find the two (indistinguishable) photons in output modes $(i,j)$ when they are injected respectively in input modes $(h,k)$, while $(P^{hk}_{ij})^{\text{D}}$ correspond to the same quantity with distinguishable particles. Note that the visibility $\mathcal{V}$ does not change in the presence of mode-dependent losses in both preparation and measurement stages, and thus its estimation gives direct access to the matrix elements. By measuring these quantities it is possible to define a system of equations and to retrieve an analytical solution to the problem, as shown in Refs. \cite{laing2012superstable,Dhand_2016}. One of the main constraints of such approach is the requirement of measurement with quantum light. Recently, a further method that exploits quantum light probes, such as two-mode squeezed states, and single-photon counting combined with heterodyne measurements, provided further refinement for reconstructing the $U$ matrix in the presence of internal unbalanced losses \cite{PhysRevA.101.043809}.

The other method proposed in Ref. \cite{Rahimi-Keshari:13}, based on classical intensity measurements, requires phase stability within the measurement time to estimate matrix phases, since these values are extracted from first-order correlation functions. This means that in such a measurement scheme the equivalence between $U$ and $U'$ does not hold. Additionally, the correct estimation of matrix element moduli, which in this approach is performed independently from the phase estimations, is spoiled by the presence of output mode-dependent losses.

\section*{Algorithm for reconstruction of linear interferometers}

In this section, we propose an alternative black-box methodology based on coherent light probes. This approach permits to estimate matrix elements moduli and losses also in the presence of loss imbalance among the modes and retrieving quantities having the same properties of $\mathcal{V}$ without requiring phase stability at the inputs and outputs of the interferometer.

\noindent
\textbf{\textit{Reconstruction of moduli and losses.}} We start our investigation with the estimation of the matrix elements moduli $\tau_{ij}$. The $\tau_{ij}^2$ coefficients represent the probability to find a single photon in output mode $j$ given an input mode $i$ or, alternatively, the fraction of classical field intensity in the mode $j$. This observation allows to define a probability matrix $P_{ij}=\tau_{ij}^2$. The main task is to estimate such a matrix when mode-dependent losses are present in the preparation and collection stages. Let us define $M$ as the matrix obtained from single-mode intensity measurements, estimated experimentally via single-photon input states or by injecting laser light in a single mode. Following the model presented in Fig.~\ref{fig:chip}a, the actual measured matrix $M$ can be expressed as follows 
\begin{equation}
    M = D_1 P D_2
    \label{eq:definition}
\end{equation}
where $D_1$ and $D_2$ are diagonal matrices that describe the unbalanced losses in input and output modes.
The task then requires reconstructing the probability matrix $P$ and the input and output losses matrices $D_1$ and $D_2$ starting from the measured matrix $M$.
To this end, we now introduce below two approaches that can be used to estimate $D_1$ and $D_2$ up to a multiplicative factor and, consequently, matrix $P$ under very few assumptions on the measured matrix $M$. 

\textit{(1) Sinkhorn's decomposition-based algorithm. --} A first approach to reconstruct matrices $P$, $D_1$ and $D_2$ is obtained starting from the observation that the probability matrix $P$ is a doubly stochastic matrix, i.e. the matrix has non-negative entries and the sum of each row and each column is equal to one. It is thus possible to apply Sinkhorn's theorem and the matrix scaling algorithm on this system \cite{Sinkhorn1967,Sinkhorn1972}.
Indeed, a matrix with non-negative elements such as $M$ admits a Sinkhorn's decomposition if it is diagonally equivalent to a doubly stochastic matrix, i.e. can be written in the form $D_1 P D_2$ where $D_1$ and $D_2$ are diagonal matrices and $P$ is a doubly stochastic matrix. This decomposition exactly represents the solution to our problem (see Eq.~\eqref{eq:definition}).
Sinkhorn's theorem and subsequent extensions \cite{Sinkhorn1967} guarantee that this solution exists and it is unique. The theorem gives us also an important warning since the algorithm is sensitive to the position of the zero elements of the measured matrix $M$. This means that an incorrect attribution of zero-valued elements in $M$, due to experimental errors or limited measurement sensitivity, could make the matrix impossible to decompose with Sinkhorn's theorem. 

Finding Sinkhorn's decomposition for a non-negative matrix $M$ is a special case of the matrix scaling problem that has applications in a large variety of fields. In our case, defining $X = D_1^{-1}$ and $Y = D_2^{-1}$, we can write $P = XMY$ and using the property that $P$ have to be doubly stochastic, we obtain the following system of equations 
\begin{equation}
    \begin{cases}
        X M Y \vec{e} = \vec{e} \\ Y M^T X \vec{e} = \vec{e}
    \end{cases}
    \label{eq:matrix scaling problem}
\end{equation}
where $\vec{e}$ is the vector with all the components equal to $1$. 

In the literature, different algorithms have been proposed to solve Eq. \eqref{eq:matrix scaling problem}. 
Here, we present a specific choice among the possible algorithms (see Ref. \cite{idel2016review} for a review of the possible approaches).
The chosen algorithm allows to recovery all the three matrices in Eq.~\eqref{eq:definition}. The idea is to rearrange Eq.~\eqref{eq:matrix scaling problem} in terms of vectors $\vec{x}$ and $\vec{y}$ which are the diagonal elements of $X$ and $Y$. Formally, they can be expressed as $\vec{x} = X\vec{e}$ and $\vec{y} = Y\vec{e}$. By defining the inversion of a vector as the inversion component-by-component, $\vec{x}^{-1} = (x_1^{-1}, x_2^{-1}, \ldots, x_{N}^{-1})^{T}$,
we find that:
\begin{equation}
    \begin{cases}
    \vec{x} = ( M \vec{y})^{-1}\\
    \vec{y} = (M^T \vec{x})^{-1}
    \end{cases}
    \label{eq:sistem_sinkhirn}
\end{equation}
At this point we apply an iterative algorithm to solve the system of equation \eqref{eq:sistem_sinkhirn} and retrieve the two vectors $\vec{x}$ and $\vec{y}$ and consequently the two diagonal matrices $X$ and $Y$. Then, we can recover the probability matrix $P = XMY$ and the two loss matrices as $D_1 = X^{-1}$ and $D_2 = Y^{-1}$.

\textit{(2) Variance-minimization based algorithm. --} 
In the derivation of the previous algorithm to solve Eq.~\eqref{eq:definition} we have implicitly assumed to have measured the field intensity distribution for any combination of outputs $j$ for any input $i$. In the following method, we define an alternative procedure that can be applied when only a subset of the inputs are available while requiring the capability of reconfiguring the linear optical network.

Let us then consider an interferometer in which it is possible to change the probability matrix $P$ without affecting the input and output losses $D_2$ and $D_1$, and to measure all the output configurations only for a subset of input modes. We first consider the scenario in which the light source, a single photon or a laser beam, is injected only in mode $i$. In absence of unbalanced losses, the outcome $\vec{M}$ of such intensity measurements is proportional to the vector $\vec{\mathcal{P}}$, (the $i$-th column of the matrix $P$) and represents a discrete probability distribution which depends from a set of parameters $\vec{\theta}$ describing the optical circuit settings. By taking into account the presence of unbalanced losses $\vec{D}$ ($D_1$ in the general case), the components $M_j$ of vector $\vec{M}$ are 
\begin{equation}
    M_j(\vec{\theta}) = I \mathcal{P}_j(\vec{\theta}) D_j
    \label{eq:variance}
\end{equation}
where $I$ is the intensity attenuated by the input loss. Since we are injecting light in the same mode for all measurements, this factor can be included in $\vec{D}$. 

The separation of the probability from the losses in Eq.~\eqref{eq:variance} can be done starting from the following observation. Let us define the quantity $S(\vec{\alpha},\vec{\theta})$ as the weighted sum of the components of $\vec{M}$: 
\begin{equation}
    S(\vec{\alpha},\vec{\theta}) = \sum_j \alpha_j M_j(\vec{\theta})
\end{equation}
If the vector of the weights $\vec{\alpha}$ is proportional to the element-wise inverse of the losses vector $\vec{D}$, the quantity $S$ does not vary with the control parameter $\vec{\theta}$. In fact, when $\vec{\alpha} = \beta \vec{D}^{-1}$, where $\beta$ is a global factor, we have:
\begin{equation}
    S(\beta \vec{D}^{-1},\vec{\theta}) = \beta \sum_j  D^{-1}_j  M_j(\vec{\theta}) = \beta \sum_j D^{-1}_j D^{\phantom{-}}_j \mathcal{P}_j(\Vec{\theta}) = \beta
\end{equation}
In general, $S(\vec{\alpha},\vec{\theta})$ changes with the controls parameters $\vec{\theta}$. Then, the idea is to find the weight vector $\vec{\alpha}$ such that $S(\vec{\alpha},\vec{\theta})$ is constant when the controls parameters $\vec{\theta}$ vary.
This vector can be obtained by minimizing the variance of $S(\vec{\alpha}, \vec{\theta})$ with respect to $\vec{\alpha}$. The variance $\xi^2(\vec{\alpha})$ can be estimated on a sufficiently large set of settings of the parameters $\{\vec{\theta_i}\}$ as
\begin{equation}
    \xi^2(\vec{\alpha}) = \frac{\sum_{i = 0}^{n-1} S(\vec{\alpha}, \vec{\theta_i})^2}{n} - \Biggl(\frac{\sum_{i = 0}^{n-1} S(\vec{\alpha}, \vec{\theta_i})}{n}\biggr)^2
\end{equation}
where $n$ is the number of different settings $\{\vec{\theta_i}\}$, and consequently represent the size of the measurement outcome vector $\{\vec{M}(\vec{\theta}_i)\}$. 
This minimization is equivalent to a quadratic optimization problem. To solve such a task, let us call $\mathcal{M}_{ij} = [\vec{M}(\vec{\theta}_i)]_j$ the matrix in which each row contain the output intensities distribution for a particular configuration of the chip then we define the following positive semi-definite matrix $Q$ as
\begin{equation}
    Q_{hk} =  \frac{1}{n} \sum_{i = 0}^{n-1} \mathcal{M}_{ih} \mathcal{M}_{ik} - \frac{1}{n^2}\sum_{i = 0}^{n-1}\sum_{j = 0}^{n-1} \mathcal{M}_{ih} \mathcal{M}_{jk}
\end{equation}
Then our problem can be rewritten, in terms of the matrix $Q$, as
\begin{equation}
    \xi^2(\vec{\alpha}) = \sum_{i}\sum_{j} Q_{ij} \alpha_i \alpha_j
    \label{eq:variance_q}
\end{equation}

The minimization of Eq. \eqref{eq:variance_q} has as trivial solution $\vec{\alpha} = \vec{0}$ (the vector with all null components) and another one that is the eigenvector of $Q$ corresponding to a null eigenvalue. The latter solution corresponds exactly to the element-wise inverse of the losses vector $\vec{\alpha} = {\vec{D}}^{-1}$. In the presence of noisy experimental data, the solution is the eigenvector of $Q$ corresponding to the lowest eigenvalue.
Alternatively, the non-trivial solution can be found by an ordinary numerical minimization approach of Eq.~\eqref{eq:variance_q}. This can be fulfilled by setting a normalization constraint $\vec{N} \cdot \vec{\alpha} = 1$ for some normalization vector $\vec{N}$, since losses can be estimated with this procedure up to a multiplicative factor common to all the modes.

The method can be generalized to the scenario in which one is interested in reconstructing a sub-matrix of $P$. 
Then, it is possible to reconstruct even the relative losses between the different measured input modes.
Here, we suppose to have a set of output vectors $\{\vec{M}_i\}_k$ for each input $k \in K$ and compute the variance function $\xi^2_k(\vec{\alpha})$ and the associated matrix $Q^{(k)}$. 
At this point we minimize the sum of all variances with the same constraints of the previous derivation: 
\begin{equation}
    \sum_{k \in K} \xi^2_k(\vec{\alpha}) = \sum_{i}\sum_{j} \sum_{k\in K} Q^{(k)}_{ij} \alpha_i \alpha_j
\end{equation}
After the minimization, it is possible to recover the input losses from the value of the sum function associated with each input as follows
\begin{equation}
    \frac{(\vec{D}_2)_k}{(\vec{D}_2)_{k'}} = \frac{\frac{1}{n}\sum_{i = 0}^{n-1} \sum_j \alpha_j \mathcal{M}_{ij}^{(k)}}{\frac{1}{m}\sum_{i = 0}^{m-1} \sum_j \alpha_j \mathcal{M}_{ij}^{(k')}}
\end{equation}

\noindent
\textbf{\textit{Reconstruction of the internal phases with classical light.}} Here we propose a procedure to estimate the complex phases of the matrix elements $\phi_{ij}$. The methods reported in Ref. \cite{laing2012superstable, Dhand_2016} employs the visibility of the Hong-Ou-Mandel (HOM) effect described in Eq.~\eqref{eq:vis} for this task, by sending a two-photon input state whose indistinguishability is tuned during the experiment. In this work, we propose an analogous quantity that can be measured by intensity cross-correlation at the output of the linear network with classical light. The measurement scheme is presented in Fig.~\ref{fig:chip}b.
The laser source is split and sent into the network in modes $(h,k)$. The additional phases $\varphi_1$ and $\varphi_2$ account for phase instabilities in the optical paths between the sources and the interferometer. We can define the cross-correlation $\sigma^{hk}_{ij}$ between the output modes $(i,j)$ when the two beams enter from modes $(h,k)$ as
\begin{equation}
    \sigma^{hk}_{ij} = \big \langle (I_i-\langle I_i\rangle)(I_j-\langle I_j\rangle)\big \rangle = \langle I_iI_j\rangle-\langle I_i\rangle\langle I_j\rangle
\end{equation}
where $I_i$ and $I_j$ are the field intensities in the corresponding output modes, while $\langle \cdot \rangle$ is the time average. We define also the self-correlation $\sigma_{ii}^{hk}$ of the intensity fluctuation as
\begin{equation}
    \sigma_{ii}^{hk} = \big \langle (I_i-\langle I_i\rangle)^2\big  \rangle = \langle I_i^2 \rangle - \langle I_i \rangle^2
\end{equation}
We make the hypotheses that (i) the external phase fluctuations $\varphi=\varphi_1 - \varphi_2$ have zero time average, and (ii) the input laser intensity is constant.
After some calculations, reported in Supplementary Materials S.II, we can define the normalized cross-correlation $C^{hk}_{ij}$ as
\begin{equation}
    C^{hk}_{ij} = \frac{\sigma^{hk}_{ij}}{\sqrt{\sigma_{ii}^{hk} \sigma_{jj}^{hk}}} = \cos(\phi_{ih}-\phi_{ik}-\phi_{jh}+\phi_{jk})
    \label{eq:cross_losses}
\end{equation}
Note that this quantity only depends on the phase of the matrix elements. Similarly to HOM visibility in two-photon experiments, the set $\{ C^{hk}_{ij} \}$ does not depend on the input and output losses. Additionally, and at variance with HOM visibility, $\{ C^{hk}_{ij} \}$ does not depend also on the moduli $\{ \tau_{i,j}\}$ of the matrix elements, thus permitting an independent estimation of the phases.
The derivation of Eq.~\eqref{eq:cross_losses} is performed under a specific assumption on the external optical phase fluctuations at the input. More specifically, we require that
\begin{equation}
    \langle e^{\imath\varphi}\rangle = 0 \qquad \langle e^{2 \imath \varphi}\rangle = 0
    \label{eq:conditions}
\end{equation}
In general, mechanical and thermal phase fluctuations do not satisfy these conditions. These equations can be satisfied by adding a phase modulator in one of the two input paths. In this scenario, the external phase contribution can be expressed as $\varphi = \varphi_M+\varphi_T$, where $\varphi_M$ is the modulated phase and $\varphi_T$ is the one modified by thermal and mechanical noise. Since the two contributions are uncorrelated, we can write $\langle e^{\imath\varphi}\rangle = \langle e^{\imath \varphi_M} \rangle  \langle e^{\imath \varphi_T}\rangle$. Controlling the phase modulation such that $\langle e^{\imath \varphi_M} \rangle = 0$ and $\langle e^{2\imath \varphi_M} \rangle = 0$, for example by adding white noise with appropriate amplitude or by a discrete set of phases, the conditions of Eq.~\eqref{eq:conditions} can be satisfied.

\noindent
\textbf{\textit{Complete algorithm.}} Given the methods defined above, we can summarize the complete procedure to reconstruct the matrix $U$ as follows
\begin{enumerate}
    \item Perform field intensity measurements in the output of the circuit. Apply the Sinkhorn-based algorithm to retrieve the complete set of moduli for the matrix elements ($\{ \tau_{i,j}\}$) or the variance minimization approach to estimate a specific subset. \label{it:algorithm1}
    \item Perform cross-correlation measurements in pairs of the output of the circuit. Solve the system of equations for the set $\{C_{ij}^{hk}\}$ to extract the complex phases of the unitary matrix. \label{it:algorithm2}
\end{enumerate}
Note that point \ref{it:algorithm2} has some similarities to the procedure proposed in Refs. \cite{laing2012superstable, Dhand_2016} with HOM visibilities. In particular, this approach could require a further minimization step on a larger set of $\{C_{ij}^{hk}\}$ with respect to the minimum ensemble needed to solve the system. Nonetheless, it is worth noting that in our case we do not require (i) the use of indistinguishable single-photon states and (ii) the measurements of $\{C_{ij}^{hk}\}$ give us directly the information about the phases without requiring the knowledge of the matrix moduli. In fact, in our algorithm points \ref{it:algorithm1} and \ref{it:algorithm2} are independent, as for the previous methods with coherent probe light \cite{Rahimi-Keshari:13, suess2020rapid}, but having the additional features of permitting the estimation of losses. Furthermore, it can be applied in any scenario with phase instabilities.

\begin{figure}[t]
    \centering
    \includegraphics[width=\columnwidth]{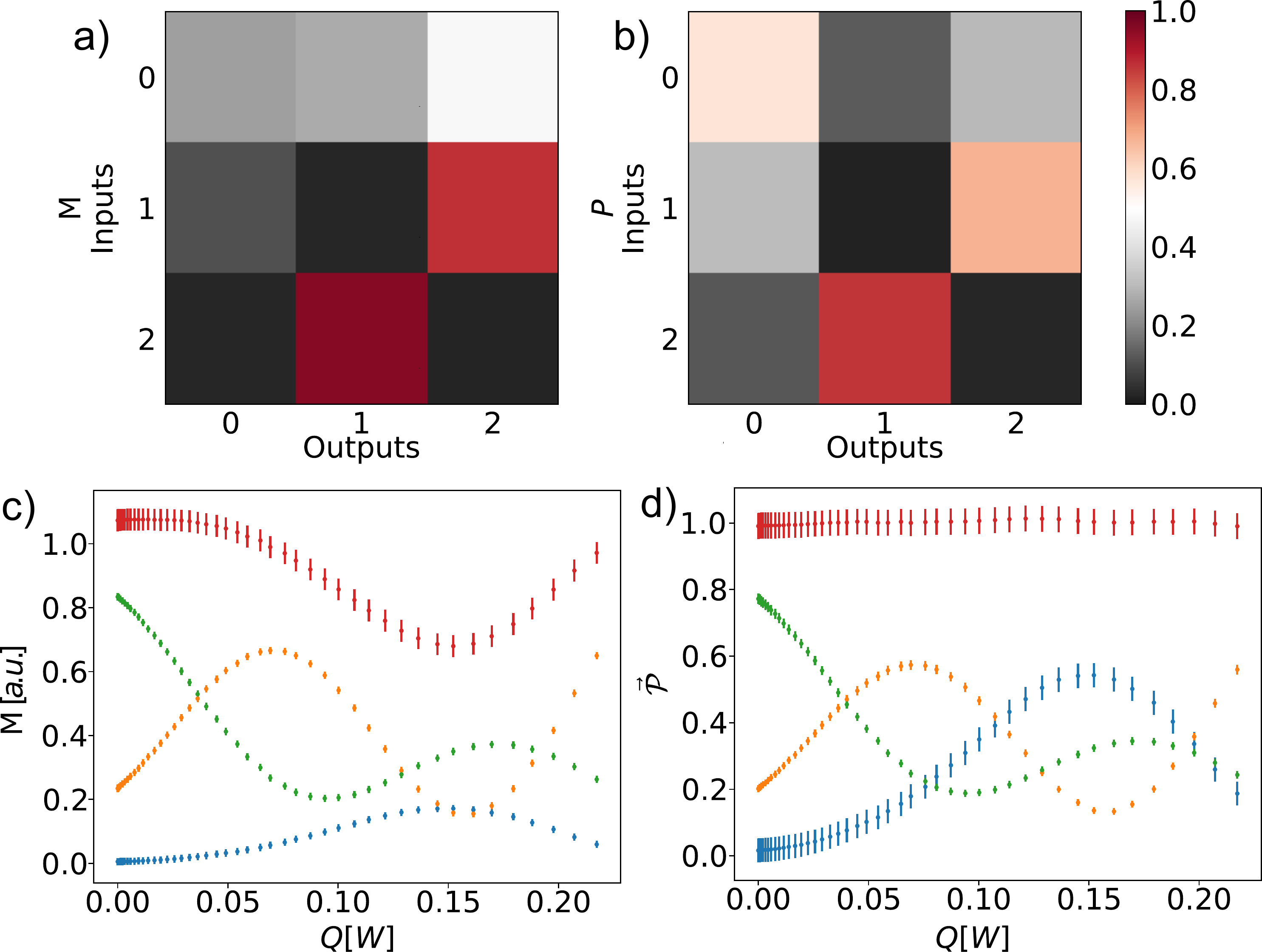}
    \caption{\textbf{Losses and moduli estimation.} We show the results of the Sinhkorn- and variance minimisation-based algorithms. First, we compare the matrix of the field intensities $M$ (panel a) with the matrix $P$ after the application of the Sinkhorn's algorithm (panel b). In figure c) we report the output intensity distribution at the three output ports when the laser is injected in the first input for different values of the electrical powers dissipated in the resistor $R_1$.  
    Red points correspond to the sum of the three intensities in the outputs (blue: output port 0, orange: output port 1, green: output port 2).
    In figure d), we report the distribution $\vec{\mathcal{P}}$ and the sum after the application of the variance minimization algorithm. The error bars reported in c) correspond to the precision of the field intensity measurements performed by a power meter. They are propagated to estimate the error of the sum. In figure d) the error bars are the result of a Monte Carlo approach applied to the reconstruction algorithm.
    }
    \label{fig:data_losses}
\end{figure}

\begin{figure*}[t]
    \centering
    \includegraphics[width=\textwidth]{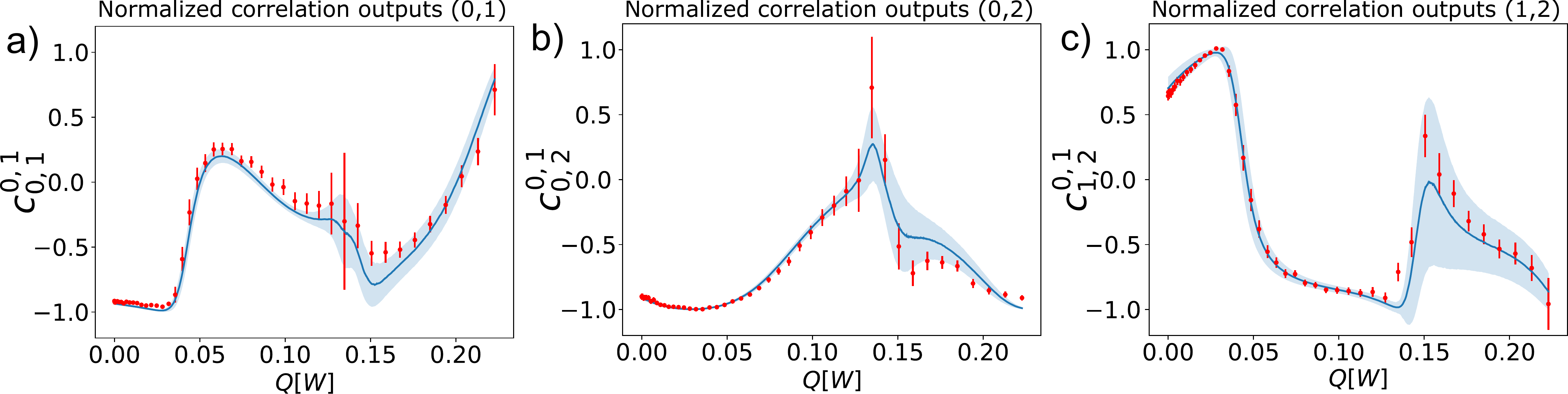}
    \caption{\textbf{Cross-correlation measurements.} In this figure we report the measurement of the normalized cross-correlations $C^{hk}_{ij}$ for different pairs of outputs, entering from $(h,k)=(0,1)$. In particular in a) we measure the pair $(0,1)$, in b) the pair $(0,2)$ and in c) the pair $(1,2)$.
In red we report the experimental correlations for different configurations of the dissipated electrical power in the heater $R_1$.
    In blue the predictions according to the results of a white-box fit that makes use of the structure of the interferometer and the previous measurements of the matrix moduli. The two independent estimations are in good agreement within one standard deviation of the experimental error.}
    \label{fig:corr_exp}
\end{figure*}

\section*{Experimental verification in a reconfigurable integrated circuit}
We tested the complete algorithm described in the previous section in a 3-mode reconfigurable optical circuit. The waveguides of the device were fabricated in a glass sample via the femtosecond laser-writing technique \cite{Gattass2008} (See Supplementary materials S.III for more details). The structure is composed of a sequence of two tritters, a circuit that generalizes the beam-splitter over three modes \cite{Spagnolo2013,Polino:19} (see also Fig.~\ref{fig:chip}c). Between the two tritters the presence of three resistive heaters permits to change the unitary transformation performed by the circuit via the thermo-optic effect. The measurements were performed with a continuous wave laser at the wavelength of $\lambda = 785 nm$. The laser is routed in different input modes via a fiber switch connected to a fiber array that injects light into the input port of the interferometer. The field intensity distribution in the 3 output ports was recorded via a CCD camera.

Our first experimental test regards the two algorithms described above, to retrieve the losses vectors and moduli of the matrix elements. In Fig~\ref{fig:data_losses} a) and b) we report the results for the application of Sinkhorn's decomposition method. In particular, in panel a) we report the measured field intensity $M$ for a particular configuration of the chip, normalized to the column total intensity. Here we inserted on-purpose additional losses by placing an attenuation filter on the output mode 0, to test the performances of the approach. In Fig~\ref{fig:data_losses} b) we report the 
probability matrix $P$ after the applications of the Sinkhorn algorithm.
We measured the moduli of a set of $N$ different transformations $U$, each of them in two loss conditions, namely by inserting and removing the attenuation filter in mode 0.
Defining the fidelity as $F = (1/3 \sum_{i,j = 0}^2 \sqrt{P_{ij} P'_{ij}})^2 $, the average classical Similarity between the two reconstructed distributions in the two losses configurations for a given $U$ via the Sinkhorn method is $\bar{F} = 0.9999 \pm 0.0001$ ($F_\text{min} = 0.9997$). The average was estimated on the set of the $N$ pairs of matrices. This confirms that the method works properly in different mode-dependent loss configurations, and retrieves as expected always the same moduli $\abs{\tau_{i,j}}^2$ associated with the transformation $U$.

We then move to test the second algorithm based on variance minimization. In Fig.~\ref{fig:data_losses} c) we report the measured field intensity in the three outputs of the interferometer for different dissipated powers in one heater, thus corresponding to different evolutions $U$, when the signal is injected in input 0. We also report the sum (red curve) of the three intensities for each tested configuration for $U$. We observe that such a sum is not constant, as one would expect if the output losses were balanced. In d) we report the same curves after applying the variance minimization algorithm, showing that the application of the algorithm makes the sum constant with respect to changes in the internal transformation.
We repeat the same procedure of the previous paragraph by tuning the output mode-dependent losses, showing the capability of retrieving the correct moduli values. 
The average fidelity, among the same set of $N$ internal operations, between the reconstructed distributions in the two different conditions of losses after the application of the algorithm is $\bar{F} = 0.9996 \pm 0.0002$ ($F_\text{min} = 0.9990$).
As a further confirmation, we compared the distributions, corresponding to the same optical circuit settings, retrieved by the application of the two algorithms. The mean fidelity between the reconstructed matrix with the two methods is $\bar{F} = 0.999 \pm 0.001$ ($F_\text{min} = 0.992$), thus confirming the effectiveness of both algorithms.

Finally, we tested the measurement of the cross-correlations defined in Eq. \eqref{eq:cross_losses} for the phase reconstruction. Since the phase fluctuations of the fibers do not fulfil the conditions of Eq. \eqref{eq:conditions}, we placed a liquid crystal in one arm before the first input of the photonic chip. For the phase modulation, we used a discrete set of phases instead of a continuous one. In particular we employed the set $\{0, 2\pi/3, 4\pi/3 \}$.
After recording the temporal fluctuations of the output field intensities, we calculated the normalized cross-correlations for various configurations of the interferometer (red dot in Fig.~\ref{fig:corr_exp}).
These measurements are compared with the predictions made by an independent characterization of the device via a white-box algorithm, used as a reference to test our approach. By this alternative method, which exploits the structure of the two tritters, the moduli and the phases of the matrix are retrieved directly from the field intensity distributions of the previous paragraph.
It follows that, in this white-box approach used as a reference, the cosines of the internal phases are the result of a numerical optimization between the parameters of the optical circuit, which is designed according to the structure in Fig.~\ref{fig:chip}c, and the measurements of the $\vec{P}$ distributions. 
The blue curve in Fig.~\ref{fig:corr_exp}) represents the prediction of such an optimization. We observe that the direct measurement of cross correlation via the proposed approach with classical light, and the same quantities retrieved via the white-box method, are compatible with the experimental error, with a normalized $\chi^2 = 1.09$.
As a final comparison, we compute the fidelity between the second column of the matrix retrieved via our black-box algorithm (bb) $U_{i1}^{\text{bb}}$ and the column of the white-box approach (wb) $U_{i1}^{\text{wb}}$. We choose the second column of the matrix since it presents non-trivial phases, namely $\phi_{i1}\ne 0$ for $i>0$. Indeed, the first row and column are imposed to be real vectors because of the equivalence between $U$ and $U'=F_1 U F_2$ with $F_1$ and $F_2$ unitary diagonal matrices. The fidelity defined here as $\mathcal{F}=\lvert \sum_{i=0}^{2} U_{i1}^{\text{bb}} (U_{i1}^{\text{wb}})^{*}\rvert^2$, i.e the overlap between pure single-photon states described by the second column of $U$, has been estimated for each configuration of the dissipated power in $R_1$ reported in Fig. \ref{fig:corr_exp}. The average fidelity is $\bar{\mathcal{F}} = 0.999\pm 0.001$ ($\mathcal{F}_\text{min} = 0.994$). 

\section*{Discussion}
In this work, we presented new algorithms to characterize the operation of multi-mode linear optical circuits. In particular, we have shown the possibility to reconstruct the moduli of the unitary matrix element by field intensity measurements, in the presence of unbalanced losses at the input and output ports. This is in contrast to the previous black-box algorithms \cite{laing2012superstable, Dhand_2016} that can reconstruct the moduli of the matrix only after phases measurements via HOM visibility and by imposing the constraint to have a unitary matrix. In addition, our procedure can provide directly the differential losses among the waveguides at the input and output. We also proposed a new method to characterize the internal matrix phases based on intensity correlations of coherent light beams at the outputs of the linear network. These measurement methods do not require knowledge of the matrix moduli and of input/output losses. These approaches enable the possibility to characterize separately the moduli and the phases of the unitary matrix implemented by the optical network, with a reliable and independent approach.
Furthermore, all presented algorithms are polynomial in the number of modes of the interferometer, both for the execution time and for the number of measurements required, sharing the same scaling of previous state-of-the-art algorithms. This property makes our algorithms suitable for adaption to large-scale interferometers.
An open issue regards the inclusion of unbalanced internal losses in the model. Some recent works have  proposed some solutions for taking into account such sources of noise in the $U$ reconstruction \cite{PhysRevA.101.043809}. We foresee as future perspective of our work to adopt the same strategy of independent estimations of moduli and phases to detect the effects of such imperfections.
We provided experimental proof of the effectiveness of the algorithm by verifying its performance on a 3-mode reconfigurable integrated optical circuit. In this analysis, we compared the results with the predictions of an independent reconstruction that exploits the knowledge of the internal structure of the circuit, which may be in general unknown or too complex to model when taking into account noise processes.

Our findings pave the way to the successful adoption of such an effective black-box methodology to large-scale optical networks, that nowadays are approaching a high number of optical modes and components \cite{Zhong_phase_2021, Hoch2021, quix_20modes}, with applications ranging from quantum communication and sensing to quantum computation and simulation via photonic systems.



\section*{Acknowledgements} 
This work is supported by the European Union's Horizon 2020 research and innovation program through the FET project PHOQUSING (``PHOtonic Quantum SamplING machine'' - Grant Agreement No. 899544), and by MIUR (Ministero dell’Istruzione, dell’Università e della Ricerca) via project PRIN 2017 “Taming complexity via QUantum Strategies a Hybrid Integrated Photonic approach” (QUSHIP) Id. 2017SRNBRK. The authors declare no competing financial interests.

\bibliography{bibliography}

\end{document}


\title{Characterization of multi-mode linear optical network \\ Supplementary materials }

\author{Francesco Hoch}
\affiliation{Dipartimento di Fisica, Sapienza Universit\`{a} di Roma,
Piazzale Aldo Moro 5, I-00185 Roma, Italy}
\author{Taira Giordani}
\affiliation{Dipartimento di Fisica, Sapienza Universit\`{a} di Roma,
Piazzale Aldo Moro 5, I-00185 Roma, Italy}
\author{Nicol\`o Spagnolo}
\affiliation{Dipartimento di Fisica, Sapienza Universit\`{a} di Roma,
Piazzale Aldo Moro 5, I-00185 Roma, Italy}
\author{Andrea Crespi}
\affiliation{Dipartimento di Fisica, Politecnico di Milano, Piazza Leonardo da Vinci, 32, I-20133 Milano, Italy}
\affiliation{Istituto di Fotonica e Nanotecnologie, Consiglio Nazionale delle Ricerche (IFN-CNR), 
Piazza Leonardo da Vinci, 32, I-20133 Milano, Italy}
\author{Roberto Osellame}
\affiliation{Istituto di Fotonica e Nanotecnologie, Consiglio Nazionale delle Ricerche (IFN-CNR), Piazza Leonardo da Vinci, 32, I-20133 Milano, Italy}
\author{Fabio Sciarrino}
\email{fabio.sciarrino@uniroma1.it}
\affiliation{Dipartimento di Fisica, Sapienza Universit\`{a} di Roma, Piazzale Aldo Moro 5, I-00185 Roma, Italy}

\maketitle

\section{Quantum correlations}

In this section, we recall the proof of Eq.~(2) of the main text.
First, we calculate the states at the outputs $i$ and $j$ when two photons, respectively in the state $\ket{\phi}$ e $\ket{\psi}$, are injected in the input modes $h$ and $k$. We first choose an orthonormal basis to express the states on the two input modes:
\begin{equation}
        \ket{\phi}_h = \sum_l \alpha_l a_{lh}^\dagger \ket{0} \quad \sum_l \abs{\alpha_l}^2 = 1 \qquad \ket{\psi}_k = \sum_m \beta_m a_{mk}^\dagger \ket{0} \quad \sum_m \abs{\beta_m}^2 = 1
    \end{equation}
Then, the two-photon states is expanded as:
\begin{equation}
        \ket{\phi}_h \otimes \ket{\psi}_k  = \sum_{l,m} \alpha_l \beta_m a_{lh}^\dagger a_{mk}^\dagger \ket{0}
\end{equation}
The state after the unitary evolution is:
\begin{equation}
    \begin{split}
        \Tilde{U} \ket{\phi}_h \otimes \ket{\psi}_k  &= \sum_{l,m} \alpha_l \beta_m \hat{U} a_{lh}^\dagger \hat{U}^\dagger \hat{U} a_{mk}^\dagger \hat{U}^\dagger \ket{0} \\
        & =  \sum_{l,m} \alpha_l \beta_m (U_{ih} a_{li}^\dagger+U_{jh} a_{lj}^\dagger)( U_{ik} a_{mi}^\dagger+ U_{jk} a_{mj}^\dagger) \ket{0}\\
        & = \sum_{l,m} \alpha_l \beta_m (U_{ih} U_{jk} a_{li}^\dagger a_{mj}^\dagger + U_{jh}U_{ik} a_{lj}^\dagger a_{mi}^\dagger + U_{ih} U_{ik} a_{li}^\dagger a_{mi}^\dagger + U_{jh} U_{jk} a_{lj}^\dagger a_{mj}^\dagger ) \ket{0}\\
        & = \sum_{l,m} \underbrace{( U_{ih} U_{jk}\alpha_l \beta_m + U_{jh}U_{ik} \alpha_m \beta_l) a_{li}^\dagger a_{mj}^\dagger}_{\text{non-collisional term}} + \underbrace{\alpha_l \beta_m (U_{ih} U_{ik} a_{il}^\dagger a_{im}^\dagger + U_{jh} U_{jk} a_{jl}^\dagger a_{jm}^\dagger)}_{\text{colisional term}} \ket{0}
    \end{split}
    \end{equation}

Since we are interested in the probability to measure two photons in two different output ports $i\neq j$, we calculate the probability that the output state is in the non-collisional term
\begin{equation}\begin{split}
        P_{ij}^{hk} &= \sum_{l,m} ( U_{ih} U_{jk}\alpha_l \beta_m + U_{jh}U_{ik} \alpha_m \beta_l)( U_{ih} U_{jk}\alpha_l \beta_m + U_{jh}U_{ik} \alpha_m \beta_l)^*\\
        &= \sum_{l,m} \abs{U_{ih}}^2\abs{ U_{jk}}^2 \abs{\alpha_i}^2 \abs{\beta_j}^2+ \abs{U_{jh}}^2\abs{U_{ik}}^2 \abs{\alpha_j}^2 \abs{\beta_i}^2+ U_{ih} U_{jk}  U_{jh}^*U_{ik}^* \alpha_l \beta_m\alpha_m^* \beta_l^* + 
        U_{jh}U_{ik}  U_{ih}^* U_{jk}^* \alpha_m \beta_l \alpha_l^* \beta_m^* \\
        &= \abs{U_{ih}}^2\abs{ U_{jk}}^2 + \abs{U_{jh}}^2\abs{U_{ik}}^2 +(U_{ih} U_{jk}  U_{jh}^*U_{ik}^*+U_{jh}U_{ik}  U_{ih}^* U_{jk}^*) \abs{\braket{\phi}{\psi}}^2
    \end{split}
    \end{equation}

In this expression, we observe that the indistinguishable photons scenario is obtained for $\abs{\braket{\phi}{\psi}}^2 = 1$, while the distinguishable particle case corresponds to $\abs{\braket{\phi}{\psi}}^2 = 0$. Using the definition of the HOM visibility, we find Eq.~(2) of the main text.
\begin{equation}
    \mathcal{V}^{hk}_{ij} = 1 - \frac{(P^{hk}_{ij})^{\text{I}}}{(P^{hk}_{ij})^{\text{D}}}= -\frac{U_{ih} U_{jk}  U_{jh}^*U_{ik}^*+U_{jh}U_{ik}  U_{ih}^* U_{jk}^*}{\abs{U_{ih}}^2\abs{ U_{jk}}^2 + \abs{U_{jh}}^2\abs{U_{ik}}^2}  = -\frac{2 \tau_{jk} \tau_{ih} \tau_{ik} \tau_{jh}} {\tau_{jk}^2 \tau_{ih}^2+ \tau_{ik}^2 \tau_{jh}^2 } \cos\left( \phi_{jk} + \phi_{ih}- \phi_{ik} -\phi_{jh}\right) 
\end{equation}

When residual distinguishability is present
the measured visibility follows the following equation:
\begin{equation}
    \mathcal{V}^{hk}_{ij} = -\frac{2 \tau_{jk} \tau_{ih} \tau_{ik} \tau_{jh}} {\tau_{jk}^2 \tau_{ih}^2+ \tau_{ik}^2 \tau_{jh}^2 } \cos\left( \phi_{jk} + \phi_{ih}- \phi_{ik} -\phi_{jh}\right) \abs{\braket{\phi}{\psi}}^2
\end{equation}
If this effect is not properly taken into account, this can lead to errors in unitary reconstruction algorithms based on such quantity.

Finally, we observe that is possible to obtain a quantity analogous to the normalized correlation of Eq.~\eqref{Eq:Normalized_correlation}, starting from visibility measurements. This procedure requires the measurements of the bunching probabilities, that is, the probability that two photons exit from the same output mode. Indeed, using this information it is possible to can reconstruct the following quantity
\begin{equation}
    {T}^{hk}_{ij} = \frac{(P^{hk}_{ij})^{\text{I}}-(P^{hk}_{ij})^{\text{D}}}{\sqrt{[(P^{hk}_{ii})^{\text{I}}-(P^{hk}_{ii})^{\text{D}}][(P^{hk}_{jj})^{\text{I}}-(P^{hk}_{jj})^{\text{D}}]}} = \frac{(U_{ih}U_{ik}^*U_{jh}^*U_{jk}+ U_{ih}^*U_{ik}U_{jh}U_{jk}^*)}{2\abs{U_{ih}}\abs{U_{ik}}\abs{U_{jh}}\abs{U_{jk}}} = \cos(\phi_{ih}-\phi_{ik}-\phi_{jh}+\phi_{jk})
\end{equation}

\section{Classical correlations}

\begin{figure}[ht!]
    \centering
    \includegraphics[scale = 0.30]{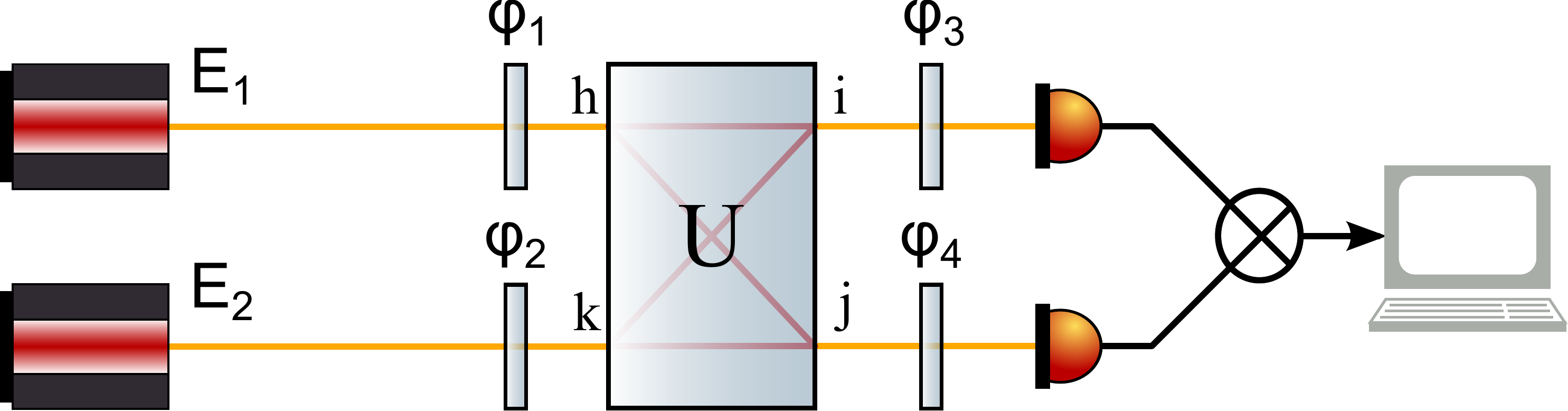}
    \caption{Conceptual scheme of the apparatus for the reconstruction of matrix phases via second-order correlations with classical light.}
    \label{fig:phase_measure}
\end{figure}

In this section, we show that second-order classical correlation measurements are described by an expression which is equivalent to two-photon Hong-Ou-Mandel visibilities.
 Starting from  Figure~\ref{fig:phase_measure}, we consider two laser beams at the input of the interferometer. The input fields are described as:
\begin{eqnarray}
    \Tilde{E}_h &=& \Tilde{E}_1 e^{i\varphi_1(t)}\\ 
    \Tilde{E}_k &=& \Tilde{E}_2 e^{i\varphi_2(t)}
\end{eqnarray}
where $\phi_1(t)$ and $\phi_2(t)$ are the phases introduced via propagation in optical fibers, and by all the other possible optical delays in the apparatus.
After propagation through the interferometer, the electric fields on the output modes read:
\begin{equation}
 \begin{aligned}
    \Tilde{E}_i &= U_{ih}\Tilde{E}_h+U_{ik}\Tilde{E}_k \\
    \Tilde{E}_j &= U_{jh}\Tilde{E}_h+U_{jk}\Tilde{E}_k 
 \end{aligned}
\end{equation}
Then, the output intensity of light is measured via two photodiodes:
\begin{equation}
 \begin{aligned}
    I_i &= \Tilde{E}_i\Tilde{E}_i^* = I_1 \tau_{ih}^2 + I_2 \tau_{ik}^2 + \Tilde{E}_1\Tilde{E}_2^*U_{ih}U_{ik}^*+\Tilde{E}_1^*\Tilde{E}_2U_{ih}^*U_{ik}\\
    I_j &= \Tilde{E}_j\Tilde{E}_j^* = I_1 \tau_{jh}^2 + I_2 \tau_{jk}^2 + \Tilde{E}_1\Tilde{E}_2^*U_{jh}U_{jk}^*+\Tilde{E}_1^*\Tilde{E}_2U_{jh}^*U_{jk}
 \end{aligned}
 \label{eq:out_intensity}
\end{equation}
If we suppose that the intensity of the input lasers $I_1$ and $I_2$ are constant and that $\langle\Tilde{E}_1\Tilde{E}_2^*\rangle = 0$ (that in our scenario is equivalent to $\langle e^{\imath(\varphi_1-\varphi_1}\rangle = 0$), we can calculate the residual as:
\begin{equation}
    \begin{aligned}
    I_i - \langle I_i \rangle &= \Tilde{E}_1\Tilde{E}_2^*U_{ih}U_{ik}^*+\Tilde{E}_1^*\Tilde{E}_2U_{ih}^*U_{ik}\\
    I_j - \langle I_j \rangle &= \Tilde{E}_1\Tilde{E}_2^*U_{jh}U_{jk}^*+\Tilde{E}_1^*\Tilde{E}_2U_{jh}^*U_{jk}
 \end{aligned}
\end{equation}
At this point we can define the cross-correlation $\sigma^{hk}_{ij}$ between the output modes $(i,j)$ when the two beams enter from modes $(h,k)$, and the self-correlation $\sigma_{ii}^{hk}$ of the intensity fluctuation a:
\begin{equation}
    \begin{aligned}
    \sigma^{hk}_{ij} &= \big \langle (I_i-\langle I_i\rangle)(I_j-\langle I_j\rangle)\big \rangle \\
    \sigma^{hk}_{ii} &= \big \langle (I_i-\langle I_i\rangle)^2\big \rangle
    \end{aligned}
\end{equation}
Let us know define a parameter $\gamma = \langle \Tilde{E}_1 \Tilde{E}_2^* \Tilde{E}_1^*\Tilde{E}_2\rangle/(I_1 I_2)$, related to the first order correlation functions of the two beams that it is in turn related to visibility of the interference fringes. In addition, we assume that the fields satisfy the following hypothesis $\langle(\Tilde{E}_1\Tilde{E}_2^*)^2\rangle = 0$, which is equivalent to the condition $\langle e^{\imath2(\varphi_1-\varphi_1}\rangle = 0$. Under these assumptions, the cross-correlations can be calculated as:
\begin{equation}
    \begin{aligned}
        \sigma^{hk}_{ij} &= \gamma I_1 I_2 (U_{ih}U_{ik}^*U_{jh}^*U_{jk}+ U_{ih}^*U_{ik}U_{jh}U_{jk}^*)\\
        \sigma^{hk}_{ii} &= 2\gamma I_1I_2 \abs{U_{ih}}^2\abs{U_{ik}}^2\\
        \sigma^{hk}_{jj} &= 2\gamma I_1I_2 \abs{U_{jh}}^2\abs{U_{jk}}^2
    \end{aligned}
\end{equation}

Finally, by defining the normalized cross-correlation $C^{hk}_{ij}$ we obtain Eq.~(19) of the main text:
\begin{equation}
    C^{hk}_{ij} = \frac{\sigma^{hk}_{ij}}{\sqrt{\sigma_{ii}^{hk} \sigma_{jj}^{hk}}} = \frac{(U_{ih}U_{ik}^*U_{jh}^*U_{jk}+ U_{ih}^*U_{ik}U_{jh}U_{jk}^*)}{2\abs{U_{ih}}\abs{U_{ik}}\abs{U_{jh}}\abs{U_{jk}}} = \cos(\phi_{ih}-\phi_{ik}-\phi_{jh}+\phi_{jk})
\label{Eq:Normalized_correlation}
\end{equation}
We note that these quantities do not depend on the visibility $\gamma$ nor from the moduli of the unitary matrix. We obtained something equivalent in the quantum correlation section but with the need for single-photon number resolving detection to evaluate the bunching probability. 

\section{Chip fabrication}
\begin{figure}[h]
    \centering
    \includegraphics[width=3cm]{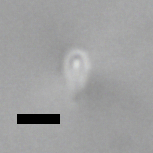}
    \caption{Typical cross-section of an optical waveguide inscribed with a single irradiation step of femtosecond laser in the borosilicate glass substrate, imaged with an optical microscope. The parameters of the laser beam and the focusing conditions are described in the text. The femtosecond laser is impinging from the top in the figure, and the scale bar corresponds to 10 $\mu$m. The refractive index modification assumes a drop-like shape with complex features; however, the waveguide supports a single optical mode at the wavelength of interest.}
    \label{fig:waveguide}
\end{figure}

The photonic chip was fabricated in a borosilicate glass substrate (EagleXG, from Corning Inc., USA)  by femtosecond laser direct writing. In detail, irradiation with focused femtosecond laser pulses causes a permanent refractive index increase in the glass, localized in the focal region (see Fig. \ref{fig:waveguide}); translation of the substrate with respect to the laser beam allows for the direct inscription of waveguides along the desired paths. 
A Yb-based cavity-dumped oscillator was adopted as a laser source, emitting ultrashort pulses with 1030~nm wavelength and 300~fs duration at the repetition rate of 1~MHz. For waveguide inscription, laser pulses of 250~nJ energy were focused 30~$\mu$m below the substrate surface by a 0.6~NA microscope objective, while the substrate was translated at a constant speed of 30~mm/s. These parameters enabled the fabrication of single-mode waveguides operating at the wavelength of 785~nm, with a $1/e^2$ mode size of $7.2 \mu m \cross 8.4 \mu m$ and propagation losses (for vertical polarization) lower than 0.8~dB/cm.
The depth of the waveguides was chosen to allow the control of the optical phase via thermo-optic phase shifters. The thermal shifters consist of metallic resistors,  manufactured by the same femtosecond laser through the ablation of a thin and uniform gold layer ($\sim 60$-nm thickness) deposited on the chip surface. Each resistor has a width of 100~$\mu$m and a length between 5 and 7 mm (parallel to the waveguide direction), which gives resistance values in the range 60-100~$\Omega$. Electrical connections are provided via standard pins, directly glued on the electrical circuit terminations.

